\documentclass[twocolumn,showpacs,preprintnumbers,amsmath,amssymb]{revtex4}
\usepackage{graphicx}
\usepackage{dcolumn}
\usepackage{bm}

\newcommand{\be}{\begin{equation}}
\newcommand{\ee}{\end{equation}}
\newcommand{\bea}{\begin{eqnarray}}
\newcommand{\eea}{\end{eqnarray}}

\newcommand{\fd}[2]{\frac{\partial #1}{\partial #2}}

\newcommand{\bra}[1]{\langle #1 |}
\newcommand{\ket}[1]{| #1 \rangle}

\newcommand{\bfr}{{\mathbf r}}

\begin{document}

\title{Pseudopotential model of ultracold atomic collisions \\
in quasi-one- and two-dimensional traps}
\author{E. L. Bolda, E. Tiesinga, and  P. S. Julienne}
\affiliation{Atomic Physics Division, 
National Institute of Standards and Technology,
100 Bureau Drive, Stop 8423, 
Gaithersburg MD 20899-8423} 
\date{\today}

\begin{abstract}
    We describe a model for $s$-wave collisions between ground state 
    atoms in optical lattices,
    considering especially the limits of quasi-one and two dimensional
    axisymmetric harmonic confinement.  When the atomic interactions are
    modelled by an $s$-wave Fermi-pseudopotential, the relative motion 
    energy eigenvalues can
    easily be obtained.  The results show that except for a bound state,
    the trap eigenvalues are consistent with one- and two- dimensional
    scattering with renormalized scattering amplitudes.  For absolute
    scattering lengths large compared with the tightest trap width, our
    model predicts a novel bound state of low energy and nearly-isotropic
    wavefunction extending on the order of the tightest trap width.

\end{abstract}    
    

\maketitle


\section{Introduction}
\label{intro}

Unprecedented control has been gained over atomic collisions through
cooling to nanoKelvin temperatures and selection of internal hyperfine
states.  Further control of these systems is now being gained by
manipulating the atoms with external fields, including magnetic fields
and optical lattices.  An opportune example where both collisions and the
external trapping potential are essential is the superfluid-Mott
insulator transition in an optical lattice.  This transition was
recently demonstrated by beginning with an atomic Bose-Einstein
condensate and adiabatically turning on an optical lattice
\cite{Greiner02}.  The final Mott insulator
state has a fixed number of atoms per lattice site.  Such a system
represents an ideal ensemble for measuring scattering properties of the
atoms.  The free-space scattering amplitudes determine the
energy eigenvalues and loss rates of the system and {\it vice versa}.

There are many current and proposed applications of cold atoms in
optical lattices.  One is high-precision measurement of atomic
potentials, by determining positions of weakly-bound molecular states
for example \cite{Bolda02a}.
Another is the implementation of quantum logic gates with neutral atoms. 
Several proposed implementations combine optical lattices with 
internal-state-dependent cold collisions \cite{Brennen99,Jaksch99}.
One can also
consider the combination of an optical lattice along one or two
directions and a weak dipole trap in the remaining directions.  Such
highly anisotropic trapping configurations were already used in
experiments on Bose-Einstein condensate number-squeezing
\cite{Orzel01} and controlled loading
\cite{HeckerDenschlag02}, as well as looking
for dimensional effects on the condensate density \cite{Goerlitz01}. 
Optical \cite{Bongs01} or magnetic \cite{Thywissen99} waveguides have
been used for guiding cold atoms.  These quasi-one- and two-
dimensional configurations can be combined with a tunable scattering
length, such as from magnetic Feshbach resonance, to lead to new
physical regimes.  Workers on Bose-Einstein condensate experiments are
observing increased phase-fluctuation in the crossover to one dimension
\cite{Dettmer01,Schreck01}.  Theoretically, the (one-dimensional)
Tonks gas regime can occur at low density when bosons are tightly confined along
two directions and weakly confined along the third
\cite{Olshanii98,Dunjko01}.  One-half anyon statistics and the
fractional quantum Hall effect may be possible for bosons confined
tightly along one direction and weakly along the other two
\cite{Paredes01}.  Another proposal is the Kosterlitz-Thouless
transition for interacting bosons in two-dimensions \cite{Stoof93}. 
For all these applications, one first needs to understand the relation
between cold collisions in free space and in a trap with arbitrary
geometry.  Then the many-body physics can be treated on the basis of
effective low-dimensional interactions.  In this article, we show how
to compute the eigenvalues from the free-space scattering data and
trap frequencies for arbitrary axisymmetric harmonic traps,
emphasizing the one- and two-dimensional regimes.  We also describe a
novel bound state induced by the trap in both low-dimensional regimes
when the scattering length is large in magnitude.

Previous theoretical work has included exact solutions of collisions for special
interatomic potentials in an {\em isotropic} harmonic trap
\cite{Busch98a,Block02}, and comparison with results including
realistic ground-state interatomic potentials.  We have
shown how an effective-scattering-length combined with a Fermi
pseudopotential can be used to calculate the energy eigenvalues for
collisions in an isotropic
harmonic trap \cite{Blume02,Bolda02a}.  We compared our model with full
interaction potential results for both a single-channel collision and
a multi-channel collision with a magnetically tunable Feshbach
resonance.  In both cases, the model can accurately treat tight traps,
as long as the trap size is larger than the van der Waals scale
length.  On the strength of this evidence we propose to also apply the
effective-scattering length model to treat ground state collisions in
axisymmetric harmonic traps. 

The effect of confinement along only one or two dimensions has also been
considered theoretically by others.  For one dimension, the scattering could be
represented by a one-dimensional delta function provided the coupling
strength was renormalized by the confining trap \cite{Olshanii98}. 
Quasi-two-dimensional scattering solutions were found for two
dimensions, where the renormalization is of the two-dimensional
scattering length \cite{Petrov00b}.  We will consider the applicability
of these two results in the case of very prolate and oblate traps
respectively.

The outline of our paper is as follows.  In the next section, we state
the problem of atoms colliding in an axisymmetric trap and the regime
of applicability of our effective-scattering-length model.  In Sec.\
\ref{selfconsistent} we show how this problem can be solved in a
particular basis.  Sec.\ \ref{quasione} contains the results of the
calculation for the one-dimensional regime and a comparison with a
one-dimensional scattering theory; Sec.\ \ref{quasitwo} contains the
same but for the two-dimensional regime.  In Sec.\ \ref{trapinduced}
we discuss a novel trap-induced bound state which appears in both the
one- and two- dimensional regimes for large positive or negative
scattering length.  We give an example of magnetically tunable
Feshbach resonance for Na atoms in quasi-one-dimension in Sec.\
\ref{example}.  We conclude the paper in Sec.\ \ref{conclusion}.  In
the appendix we derive the matrix elements of the anisotropic
potential required for the calculation.

\section{Two atoms colliding in an anisotropic harmonic trap}
\label{twoatoms}

We assume ultracold atoms are trapped in an optical lattice detuned far off 
resonance.
Specific optical lattice potentials for different angular and polarization
configurations are calculated in \cite{Petsas94}.  For our purposes we assume
that two atoms in specific internal sublevels
remain near a local minimum of the potential.  With the assumption of local
azimuthal symmetry about an axis through a potential minimum, we approximate
the anisotropic potential near a particular site by
\bea
V_t(\bfr_j) &  = & \frac{1}{2} m \left[ \omega_\bot^2 ( x_j^2 + y_j^2) + 
\omega_z^2  z_j^2 \right] \,,
\eea
where $ \bfr_j$ is the positions of atom $j=1$ or $2$,  $m$ is the atomic
mass, and $\omega_\bot$ and $\omega_z$ are trapping frequencies.
We define the trap anisotropy
\be
 A = \frac{\omega_z}{\omega_\bot} 
\ee
so that one-dimensional physics is approached for the oblate trap $A \ll 
1$ ("cigar") and two-dimensional physics for the prolate trap $A \gg 1$("pancake"). 
The  length scales associated with the transverse and longitudinal trap 
directions are
\be
d_\bot = \sqrt{\frac{ \hbar}{\mu \omega_\bot}} \,\,,\quad
d_z = \sqrt{\frac{ \hbar}{\mu \omega_z}} \,,
\ee
where $\mu=m/2$ is the reduced mass of the atom pair.

As in the case of an isotropic harmonic trap, the center
 of mass and relative motion are separable.  The center of mass motion
 is independent of the interatomic potential, $V_{\mathrm{int}}(r)$, and has the usual harmonic
 oscillator solutions.  The relative coordinate  Hamiltonian is given in spherical
 coordinates by
 \be
 {\hat H} = -\frac{\hbar^2}{2 \mu r^2} \fd{}{r} \left( r^2 \fd{}{r} 
 \right) + \frac{{\hat L}^2}{2 \mu r^2} + V(r,\theta) + V_{\mathrm{int}}(r)
 \label{Hamiltonianspher}
 \ee
 where $\bfr = \bfr_1 - \bfr_2$ and $r = |\bfr|$. The interatomic 
 orbital angular momentum operator${\hat L}$ gives the partial wave quantum 
numbers $l=$ 0, 1, 2, ... for $s$-, $p$-, $d$-, ... waves.
 The potential due to the trap written in terms of the spherical 
 harmonic $Y_{20}$ is
 \be
V(r, \theta) =  \frac{1}{2} \mu \omega^2 r^2 \left[ 1 +  \sqrt{\frac{16 
\pi}{5}} \Lambda Y_{20} 
 (\theta, 0) \right] \,,
\label{TrapV}
 \ee
 where the mean-square trap frequency appears as
 \be
 \omega = \sqrt{ \frac{2 \omega_\bot^2 + \omega_z^2}{3}}
 \ee
  and
 \be
 \Lambda =  \frac{\omega_z^2 - 
 \omega_\bot^2}{\omega_z^2 + 2 \omega_\bot^2} =  \frac{A^2 - 1}{A^2 + 2} \,.
 \label{Lambdadef}
 \ee
The term proportional to $\Lambda$ in Eq.~(\ref{TrapV}) defines the 
 anisotropic part ${\hat H}^{(1)}$
 of the Hamiltonian ${\hat H} = {\hat H}^{(0)} + {\hat H}^{(1)}$.  The length scale associated with the mean-square frequency is defined to be
 \be
 d = \sqrt{\frac{\hbar}{\mu \omega}} \,.
 \ee

The interatomic $V_{\mathrm{int}}(r)$ potential for two ground state
atoms approaches $-C_6/r^6$ at large internuclear separation $r$.  The
associated van der Waals length scale is $x_0 = (2\mu
C_6/\hbar^2)^{1/4}/2$~\cite{Gribakin93,Weiner99,Williams99}.  It gives
the approximate size of the potential, that is, the wavefunction takes
on its asymptotic scattering form for $r \gg x_0$.  We have previously
approximated the exact Born-Oppenheimer potential by the
energy-dependent Fermi pseudopotential \cite{Bolda02a,Blume02,Huang57}
\be
{\hat V}_{\mathrm{eff}}(r; E) = \frac{4 \pi \hbar^2 a_{\mathrm{ eff}}(E)}{m}
\delta(\bfr) \fd{}{r}r \,.
\ee
The dependence on 
collision energy $E$ is due to the energy-dependent scattering length, 
defined as
\be
 a_{\mathrm{eff}}(E) = -\frac{\tan \delta_0(E)}{k} \,,
\label{aeff}
\ee
where $E=\hbar^2 k^2/2 \mu$ and $\delta_0(E)$ is the $s$-wave 
collisional phase shift.  $S$-wave scattering predominates for ultracold collisions except for the case of identical fermions.

The pseudopotential approximation is valid provided the  van der Waals 
length scale $x_0$ is less than the smallest harmonic oscillator width, 
$x_0 \ll \mbox{min}\{ d_\bot, d_z \}$~\cite{Bolda02a}.   We assume 
that the energy shifts due to higher partial waves are negligible in comparision 
to those of the $s$-wave.  This tends to be true for ultracold collisions 
because the centrifugal barrier heights are large compared to the collision 
energy.   We also neglect inelastic losses, that is, the imaginary part of the 
scattering length is much smaller than the real part \cite{Tiesinga00a}.
 
Since spherical symmetry is
broken by the anisotropic potential, $\hat{L}^2$ does not commute with 
the Hamiltonian.  Consequently, partial waves with the same parity with 
respect to $l$ are coupled.   While the projection $\hat{L}_z$ of angular
 momentum on the $z$-axis  does commute with the Hamiltonian,
only its $m_l=0$ eigenstates are affected by $s$-wave scattering.  Thus 
we only compute the energies of even partial wave, $m_l=0$ states.  We do 
not consider odd partial waves, should they be present for distinguishable 
bosons, because they have negligible energy shifts in the limit of very low 
collision energy.

\section{Method  of eigenvalue solution and self-consistent energies}
\label{selfconsistent}

We need to solve the eigenvalue problem for the Hamiltonian Eq. 
(\ref{Hamiltonianspher}) self-consistently, because of the
energy-dependent scattering length in the pseudopotential term.  
As in Ref.~\cite{Bolda02a}, this is done in two steps. We
first obtain the eigenvalues $E_i(a/d,A)/\hbar \omega$ in scaled trap
energy units for fixed values of $A$ and the energy-{\em independent}
scaled scattering length $a/d$.  The self-consistent energy
eigenvalues for an actual system with $a_{\mathrm{eff}}(E)$ from
Eq.~(\ref{aeff}) are then found graphically, for a given $A$, by
superposing a plot of $E_i(a/d,A)$ as a function of $a/d$ and a plot
of ${a_{\mathrm{eff}}(E) }/{d}$, with $E$ as the ordinate and
$a_{\mathrm{eff}}/d$ as the abscissa (cf. Fig.~1 and Fig~2 of \cite{Bolda02a}).  The points where the curves intersect determine the self-consistent energies.

In the rest of this section we focus on obtaining $E_i(a/d,A)$, since $a_{\mathrm{eff}}(E)$  can be found from a standard free-space scattering calculation.  We
use the partial wave expansion of the wavefunction,and expand each
partial wave, with the exception of $s$-waves, in {\em isotropic}
harmonic oscillator eigenfunctions of frequency $\omega$.  For the
$s$-wave part, we use the analytic eigenfunctions of the isotropic
harmonic oscillator with a Fermi pseudopotential proportional to
scattering length $a$ \cite{Busch98a}.  These automatically
incorporate the singular nature of the wavefunction at the origin.  Since $m_l=0$ we can set the spherical coordinate $\phi = 0$, and consequently we write
 \bea
\psi(r, \theta; a) &=& \sum_n c_{n 0} Q_n(r; a) Y_{00}(\theta,0) \nonumber \\
&& \quad + \sum_{n} \sum_{l>0} c_{n l} R_{nl}(r) 
Y_{l0}(\theta,0)
\label{expand}
\eea
\be
Q_n(r; a) =  \frac{2 a}{\sqrt{\pi}d^2} 
\sqrt{\fd{\nu_n}{a}} \Gamma(-\nu_n) U \left(-\nu_n, \frac{3}{2}; 
\frac{r^2}{d^2} \right) e^{-\frac{r^2}{2 d^2}}
\label{Qdef}
\ee
\be
R_{nl}(r) = \left[ \frac{2 (n!)}{\Gamma(n+l+\frac{3}{2})} 
\right]^{\frac{1}{2}}
\left(\frac{r}{d} \right)^{\frac{l}{2} + \frac{3}{4}} 
L_n^{(l+\frac{1}{2})} \left( \frac{r^2}{d^2} \right) e^{-\frac{r^2}{2 
d^2}} 
\label{Rnldef}
\ee
where the summations are over all nonnegative integers $n$ and even 
$l>0$.
Here $U$ is the (second) Kummer confluent hypergeometric function,
$L_n^{(l+\frac{1}{2})}$ 
are the Laguerre polynomials, and $\Gamma$ is the Gamma function 
\cite{Abramowitz72}. The nonintegral $s$-wave 
quantum numbers $\nu_n$ for an isotropic trap are determined by
\be
\frac{a}{d} = \frac{1}{2} \tan \pi \nu_n \frac{\Gamma(\nu_n + 
1)}{\Gamma(\nu_n + 
\frac{3}{2})}.
\ee
This equation also is used in the calculation of the derivative in Eq. 
(\ref{Qdef}).   
 
We need the Hamiltonian matrix elements in the basis used in Eq.~(\ref{expand}). The isotropic part of the Hamiltonian Eq.  (\ref{Hamiltonianspher}) gives only diagonal matrix elements:
\be
H^{(0)}_{nl;nl}=\left(2 n + l + \frac{3}{2}\right) \hbar \omega
\ee
for $l>0$.
Only the $s$-wave diagonal matrix elements,
\be
H^{(0)}_{n0;n0}=\left(2 \nu_n + \frac{3}{2}\right) \hbar \omega \, ,
\ee
 are affected at low energy by atom-atom 
interactions proportional to the scattering length.   The
 anisotropic part of the Hamiltonian, Eq.  (\ref{Hamiltonianspher}) contributes both 
diagonal and off-diagonal matrix elements:
 \be
 H^{(1)}_{nl;n'l'} = \sqrt{\frac{4 \pi}{5}} \mu \omega^2 \Lambda \bra{n, l}  r^2  Y_{20}(\theta,0) 
 \ket{n', l'}
 \ee
 for all principal quantum numbers $n,n'$ and even partial waves $l,l'$. 
 The derivation of these matrix elements is given in the Appendix.  The
 diagonalization of the Hamiltonian matrix is straightforward with sparse
 matrix eigenvalue routines.  For the most extreme anisotropies 
 considered in this 
 paper (as small as $0.01$ and as 
 large as $100$), we required a maximum $l = 600$ and a maximum 
 $n=600$ to compute the lowest few eigenvalues for all values of scattering 
 length.  We have checked that the correct solutions are approached 
 as $A \rightarrow 1$.

\section{Quasi-one-dimensional trap }
\label{quasione}

We have found solutions in the quasi-one-dimensional regime $A \ll 1$. 
Figure~\ref{A001vsa} shows $E_i(a/d,A=0.01)$ versus $a/d$.  The figure
also shows the eigenvalues $E_i^{1D}(a/d)$ for a purely one-dimensional
model, corresponding to interaction via a delta function in $z$ and 
trapping along $z$ only. Following Ref.~\cite{Busch98a}, the eigenvalues are 
\be
E_{i}^{{1D}} = \left(\frac{3}{2}\right)^{\frac{1}{2}}(1 + 
\frac{A^{2}}{2})^{-\frac{1}{2}} \left[1 + A \left(2 \nu_{i}^{1D} + 
\frac{1}{2} \right)
\right] \hbar \omega
\ee
where the $\nu_{i}^{1D}$ satisfy
\be
\tan \pi \nu_{i}^{1D} \, \frac{\Gamma (\nu_{i}^{1D} + 1)}{\Gamma 
(\nu_{i}^{1D} + 
\frac{1}{2} )} = g^{1D}  \, .
\ee
The one-dimensional interaction, $\hbar \omega g^{1\mathrm{D}} \delta(z/d)$,
 is related to the three-dimensional scattering length through \cite{Olshanii98} 
\be
g^{1D} = \frac{\left(\frac{3}{2} \right)^{\frac{1}{4}}A^{-\frac{1}{2}}(1 + 
\frac{A^{2}}{2})^{-\frac{1}{4}}\frac{a}{d}}
{1 - 1.4603 \left(\frac{3}{2} \right)^{\frac{1}{4}}(1 + 
\frac{A^{2}}{2})^{-\frac{1}{4}} \frac{a}{d}} .
\label{g1d}
\ee

Our eigenvalues agree well with those from
the one-dimensional model for $E>0.8 \hbar \omega$.  To get this 
agreement, it is crucial to include the renormalization in the denominator 
of Eq.~(\ref{g1d}).  The levels for  $E > 1.23 \hbar \omega$ represent trap levels aligned along the weak trapping direction with spacing about $2 \hbar \omega_z = 0.024 \hbar \omega$.  In Fig. ~\ref{A001vsa}, there is no difference within our numerical accuracy between our lowest three trap levels and those from the one-dimensional model maximum energy difference, while the difference for the highest trap level shown is $0.002 \hbar \omega$.  The eigenvalues 
for large positive or negative scatttering length
approach the same asymptotic values. The lowest energy state is 
not predicted accurately by the one-dimensional
model; we discuss this special state further in Sec.\ \ref{trapinduced}.

\begin{figure}
    \caption{Energy eigenvalues versus scattering length at $A = 0.01$ from Fermi-pseudopotential 
    (solid line) and one-dimensional scattering 
    theory (dashed line).($\nu = \omega/2 \pi$).}
 \label{A001vsa}
 \end{figure}

The wavefunction $r \psi(x=r \sin{\theta}, y=0, z= r\cos{\theta})$ , corresponding to the
second lowest energy, with $A=0.01$ and $a/d = -25$, is plotted in Fig.~\ref{secondwavefn0.01}. 
In Fig.~\ref{secondwavefn0.01}(a) the one
dimensional nature of the wavefunction is apparent on a scale large 
compared with $d$.  The variation in
$x$ is approximately Gaussian, while scattering results in the dip
along the line $z = 0$.  A close-up of the wavefunction in 
Fig.~\ref{secondwavefn0.01}(b) reveals how the scattering crosses over to a
isotropic three-dimensional character at short interatomic distance.  The function
$r\psi$ is finite at the origin as a consequence of the pseudopotential scattering.

\begin{figure}
      \caption{Wavefunction $r \psi(x,0,z)$ in Cartesian 
      coordinates $x=r \sin{\theta}, y=0, z= r\cos{\theta}$, corresponding to second lowest energy eigenvalue for 
      $A = 0.01$ and $a/d = -25$. (b) is a close up of (a).  All lengths are expressed in trap units of $d$ and $\psi$ in units of $d^{-3/2}$.}
    \label{secondwavefn0.01}
 \end{figure}

We have compared some of our results to a recent diffusion quantum Monte-Carlo 
study on the ground state of interacting bosons in elongated traps 
\cite{Blume02b}. Our lowest trap state agrees with the ground state 
of that method to within one percent for 
anisotropies in the range $1 \ge A \ge 0.01$ for a {\it fixed} positive value of 
the scattering length, even though that study assumed a hard-core 
potential of size $a$.

\section{Quasi-two-dimensional trap}
\label{quasitwo}
We have also computed solutions in the quasi-two-dimensional regime, $A \gg 1$. Figure~\ref{A100vsa} shows
$E_i(a/d,A=100)$ versus $a/d$.  The figure also shows the eigenvalues
$E_i^{2D}(a/d)$ for a purely two-dimensional model, corresponding to
interaction via zero-range two-dimensional scattering and trapping in the 
$x,y$-plane only.
Again following \cite{Busch98a}, the eigenvalues of a two-dimensional trap are
\be
E_{i}^{2D} = \left(\frac{3}{2}\right)^{\frac{1}{2}}(1 + 
\frac{A^{2}}{2})^{-\frac{1}{2}} 
\left[\frac{A}{2} + 2 \nu_{i}^{2D} + 1\right] \hbar \omega \, ,
\ee
where the $\nu_{i}^{2D}$ satisfy
\be
\Psi(-\nu_{i}^{2D}) = \frac{1}{g^{2D}}
\ee
and $\Psi$ is the digamma function \cite{Abramowitz72}. 
The two-dimensional scattering is mediated by an 
interaction strength related to the three-dimensional scattering length through
\be
g^{2D} =  \frac{
    \left(\frac{3}{2\pi^2}\right)^{\frac{1}{4}} A^{\frac{1}{2}} \left( 1 +\frac{A^2}{2}  
\right)^{-\frac{1}{4}} \frac{a}{d}       }
{1+  \left(\frac{3}{2\pi^2}\right)^{\frac{1}{4}}\ln \left(\frac{0.915A}{4\pi}\right)
     A^{\frac{1}{2}} \left( 1+ \frac{A^2}{2} 
\right)^{-\frac{1}{4}} \frac{a}{d} 
} \, .
\label{g2dXX}
\ee 
This expression for $g^{2D}$ is derived by simple algebra from equations 
in Ref.~\cite{Petrov00b}.

\begin{figure}
    \caption{Energy eigenvalues versus scattering length at $A = 100$ from Fermi-pseudopotential 
    (solid line) and two-dimensional scattering 
    theory (dashed line).($\nu = \omega/2 \pi$)}
 \label{A100vsa}
 \end{figure}

Our eigenvalues agree well with this model for $E>0.8 \hbar \omega$.  To obtain this agreement it is crucial to include the renormalization in the denominator of Eq. (\ref{g2dXX}). The trap levels above $E \approx 0.9 \hbar \omega$ are spaced by about $2 \hbar \omega_\bot = 0.03462 \hbar \omega_\bot$.   In Fig. ~\ref{A100vsa}, the difference between our lowest trap level and that from the two-dimensional model is $0.0015 \hbar \omega$ at $a/d = \pm 20$; the difference for the highest trap level shown is $0.002 \hbar \omega$ at the same values of scattering length.
The eigenvalues for large positive or negative scattering length
approach the same asymptotic values. Again we note the exception that the lowest energy state is not predicted well by the renormalized two-dimensional model (see 
Sec.~\ref{trapinduced}).  A feature of the two-dimensional physics seen in
Fig.~\ref{A100vsa} is that for $|a| \ll d$, all eigenvalue curves
except the lowest have nearly the same slope.

A sample wavefunction $r \psi(x=r \sin{\theta}, y=0, z=r \cos{\theta})$ , corresponding to the 
second lowest energy, at $A=100$ and $a/d = -25$, is plotted in Fig.~\ref{secondwavefn100}.  (Note that the $x$ and $z$ axes are interchanged from Fig. \ref{secondwavefn0.01}.)
\begin{figure}
      \caption{Wavefunction $r \psi$ in Cartesian 
      coordinates $x=r \sin{\theta},y=0,z=r \cos{\theta}$, corresponding to second lowest energy eigenvalue for 
      $A = 100$ and $a/d = -25$. (b) is a close up of (a).  All units are as in 
Fig.~\ref{secondwavefn0.01}.}
    \label{secondwavefn100}
 \end{figure}
The two-dimensional nature of the physics is immediately apparent  on a scale large compared with $d$. The
variation in $z$ is approximately Gaussian, with a dip along $x = 0$. 
The close-up Fig.~\ref{secondwavefn100}(b) reveals how the scattering crosses over to a
isotropic three-dimensional character at short interatomic distance, as in the quasi-one-dimensional case. 

\section{Trap-induced bound state}
\label{trapinduced}

Curiously, for large magnitudes of the scattering length, the lowest
energy eigenvalue appears near $\hbar \omega/2$ for all values of
trap anisotropy.  This behavior can be explained by examining the
wavefunctions of these states plotted in
Figs.~\ref{lowestwavefn0.01} and \ref{lowestwavefn100}, at $a = -25$,
for $A = 0.01$ and $100$ respectively.  
\begin{figure}
 \caption{Wavefunction $r \psi$ in Cartesian 
      coordinates $x=r \sin{\theta},y=0,z=r \cos{\theta}$, corresponding to lowest energy eigenvalue for  $A = 0.01$ and $a/d = -25$.  All units are as in 
Fig.~\ref{secondwavefn0.01}.}
    \label{lowestwavefn0.01}
 \end{figure}
\begin{figure}
           \caption{Wavefunction $r \psi$ in Cartesian 
      coordinates $x=r \sin{\theta},y=0,z=r \cos{\theta}$, corresponding to lowest energy eigenvalue for  $A = 100$ and $a/d = -25$.  All units are as in 
Fig.~\ref{secondwavefn0.01}.}
    \label{lowestwavefn100}
 \end{figure}
Both wavefunctions
are more nearly isotropic than those corresponding to higher levels at 
the same parameters (see Figs.~\ref{secondwavefn0.01} and \ref{secondwavefn100}), 
and their extent is roughly the mean trap length scale $d$. 
Similar wavefunctions are obtained for large positive scattering
length.

The energy of the lowest state is plotted in Fig.~\ref{lowesteigvsA}
for $a/d = \pm 100$ as a function of anisotropy.  
    \begin{figure}
      \caption{Lowest-energy eigenvalue versus trap anisotropy $A$ (logarithmic scale) at $a/d = 
      -100$ (triangles) and  $a/d = 100$ (squares). Fits up to first order in 
      $\Lambda^2$ for $a/d = -100$ (solid line) and $a/d = 
      100$ (dashed line) are shown; see text for coefficients of fits. The fits verify that the wavefunction is nearly isotropic for all $A$. ($\nu=\omega/2\pi$)}
     \label{lowesteigvsA}
  \end{figure}
The shape of this
curve can be understood from a perturbative picture in the anisotropic interaction
$H^{(1)}$, defined in Eq.\ (\ref{Lambdadef}).
To zeroth-order in this picture the state is  the
lowest level of the {\em isotropic} trap of frequency $\omega$
and a pseudopotential with scattering length $a$.
(Recall that for an isotropic trap with $a \rightarrow -\infty$ the lowest energy 
is  $\hbar \omega/2$.) There is no first-order correction in $H^{(1)}$,
while to second-order the energy gets a small correction proportional to
$\Lambda^2$.
Indeed, we find that a quadratic fit in $\Lambda$
at $a/d = -100$ results in $E_0/\hbar\omega = 0.5054 - 0.0562 \Lambda^2$;
compare with the zeroth-order energy $0.5056 \hbar\omega$ for the lowest state in an isotropic
trap at the same scattering length.  Similarly, at $ a/d = 100$ we
find $E_0/\hbar\omega = 0.4941-0.0549 \Lambda^2$ compared with the zeroth-order energy $0.4943\hbar\omega$ in the isotropic case.  

When neglecting the pseudopotential,
the energy of the lowest  state is $E(a=0)/\hbar \omega = (1 +
A/2)/\sqrt{2/3 + A^2/3}> \sqrt{3}/2$, so we should properly
consider a state to be bound when its energy is lower than
this.  Thus we denote such a state the {\em trap-induced bound
state}.  One can also think of it as an artificial molecule with an
extent given roughly by the size of the tightest trap dimension.  As
the scattering length approaches zero from the positive side, we
recover the usual  molecular bound state with an energy below zero.

\section{Feshbach resonance in quasi-one-dimensional trap}
\label{example} 
One possible way of varying the atomic interaction strengths experimentally 
is through the use of a tunable Feshbach resonance state.   Consequently we 
describe a quasi-one-dimensional
magnetically tunable Feshbach resonance by using the self-consistent 
energy method with the eigenvalues of Fig.~\ref{A001vsa}.  We consider 
two Na atoms in
their lowest hyperfine levels, for which an $s$-wave Feshbach
resonance occurs near 90.9 mT \cite{Inouye98,vanAbeelen99,Mies00}. 
The scattering length is highly energy- and magnetic-field- dependent near the resonance.
We use the effective scattering length from a close-coupling calculation, 
as described in Ref.~\cite{Bolda02a,Mies00}.
The trap frequencies are taken to be $\omega_\bot/2 \pi = 612$ kHz,
$\omega_z/2 \pi = 6.12$ kHz, so the trap anisotropy is $A = 0.01$ and
$\omega/2 \pi = 500$ kHz.  Using the procedure outlined at the
beginning of Sec.\ \ref{selfconsistent}, we predict the eigenvalues
as a function of applied magnetic field near the resonance in
Fig.~\ref{EvsB0.01}.  
As the magnetic field is tuned through
resonance, the lowest state goes continuously from a molecular state
with $E < 0$, to the trap-induced bound state $E \approx \hbar
\omega/2$, to the lowest quasi-one-dimensional trap state at $E
\approx \sqrt{3/2}\hbar\omega \approx 1.23 \hbar \omega$. For the trap frequencies
chosen here one can change
from a molecular bound state to a trap state by tuning the magnetic field 0.01 mT.
The trap states are smoothly shifted up by $2 \hbar\omega_z$ as the magnetic field is increased.

\begin{figure}
      \caption{Energy eigenvalues versus magnetic field $B$ for two Na atoms in the lowest 
      hyperfine level in a axisymmetric trap with $\nu = \omega/2 \pi = 500$kHz and 
      anisotropy $A = 0.01$. The dashed  line shows the energy of the lowest trap 
      level when the interatomic interaction is neglected.}
    \label{EvsB0.01}
 \end{figure} 

\section{Conclusion}
\label{conclusion}

We have argued that an energy-dependent pseudopotential approach may
be used to calculate eigenvalues of two ultracold atoms colliding in
an axisymmetric harmonic trap.  Furthermore, we have numerically
solved for the eigenvalues of the axisymmetric trap with an $s$-wave
pseudopotential interaction proportional to a scattering length. 
These results can be considered a generalization of the isotropic trap
case previously solved \cite{Busch98a,Bolda02a}.  Our results show
that one- and two- dimensional interaction regimes can be reached, but
that the interactions become renormalized by the tight trapping
potential when the magnitude of the effective scattering length is
large compared with a mean trap length.  Remarkably, in the case of
scattering length of large magnitude, we find a nearly isotropic state
with energy near $\hbar \omega/2$ for all values of trap anisotropy. 
This is a trap-induced bound state. The size of state is controllable
by the tightest trap frequency.  We show by an example that this state
can be reached with the current techniques of magnetically tunable
Feshbach resonance applied to atoms in an optical lattice.

The numerical techniques used in this article may also be useful
when the interatomic interaction becomes sufficiently
long-range compared with the tightest trap direction or is anisotropic such that the
$s$-wave pseudopotential approach becomes insufficient. 
This is of importance, for example, for dipole-dipole interactions \cite{Brennen99,Goral02,Derev03}.  The possibility of
manipulating shape resonances (such as the $d$-wave resonance in Na
collisions) with the trap should not be overlooked.  The
effective-scattering-length model and partial wave expansion in our
numerical technique should
also solve the eigenvalue problem for atoms colliding in separated traps, as can occur
in a state-dependent optical lattice.   This is particularly important for the
proposals on quantum computing with neutral atoms.  We are continuing
work in this area.


\section*{Acknowledgments}
Discussions with M. Olshanii and B. Gao were helpful. We thank D. 
Blume for making available her results on two-body ground state energies.
ELB was supported from the National Research Council. 
ET and PSJ acknowledge support from the Office of Naval Research.
 
\section*{Appendix: Matrix elements of anisotropic potential}
\label{appendix}
 
In this appendix we evaluate the matrix elements of the anisotropic
potential term $H^{(1)}_{nl;n'l'}$ in the basis of partial waves,
isotropic harmonic oscillator functions and the irregular $s$-wave
oscillator eigenfunctions. 
 
The partial wave expansion applied to the angle-dependent factor of the 
Hamiltonian is evaluated with the three-spherical-harmonic 
formula,
\bea
I_{ll'} & = &  \sqrt{\frac{4 \pi}{5}} \int d \Omega \, Y_{l'0}^*(\Omega) Y_{20}(\Omega) 
Y_{l0}(\Omega)  \nonumber \\
& = & \sqrt{\frac{2 l + 1}{2 l'+1}} 
 \langle 2 l ; 0 0 | 2 l ; l' 0 \rangle^2.
\eea
Evaluation of the Clebsch-Gordan coefficient $\langle 2 l ; 0 0 | 2 l ; l' 0 
\rangle$ shows that
\bea
I_{ll} & = &  \frac{l(l+1)}{(2 l - 3)(2 l +3)} \\
I_{l,l+2} & = & I_{l+2,l} =  
\frac{3(l+1)(l+2)}{2 (2 l +3) \sqrt{(2 l+1)(2 l+5)}}
\eea
while all other angular matrix elements are zero.

This leaves the radial factor of the matrix element to be computed. 
Throughout the remainder of the appendix kets refer to the radial part
of the basis functions only, so that for (cf.  Eqs.  (\ref{Qdef}) and
(\ref{Rnldef}))
\bea
\langle r \ket{n l} & = & R_{nl}(r) \\
\langle r \ket{n0} & = & Q_n(r;a) \,.
\eea
In this notation
\be
H^{(1)}_{nl;n'l'} = \mu \omega^2 \Lambda I_{ll'} \bra{nl} r^2 \ket{n'l'} \,.
\ee

The analytic evaluation of the radial matrix elements $\bra{n l} r^2
\ket{n'l}$ is most conveniently carried out using $n$- and $l$-ladder
operators.  From the factorization method applied to the radial
Schr\"{o}dinger equation \cite{Infeld51}, the $n$-raising and
$n$-lowering operators are
\be
\hat{b}_{nl}^{\pm} = \pm r \fd{}{r} \pm \frac{1}{2} - r^2 + 2 n + l + 
\frac{1}{2} 
\ee

\bea
\hat{b}_{nl}^{-} \ket{ n l} & = & \sqrt{2 n (2 n + 2 l +1)} \ket{n-1, 
l} \\
\hat{b}_{n l}^{+} \ket{ n-1, l} & = & \sqrt{2 n (2 n + 2 l +1)} \ket{n l},
\eea
while the $l$-raising operator is
\be
\hat{\cal L}_l^+ = \fd{}{r} + r - \frac{l}{r}
\ee
such that
\be
\hat{\cal L}_l^+ \ket{nl} = - 2 \sqrt{n} \ket{n-1, l+1}.
\ee
(Note that these operators are applied to the normalized 
radial eigenfunctions.) 

For $l = l'>0$, we use the fact that
\be
r^2 = 2 n + l + \frac{3}{2} - \frac{1}{2} \hat{b}_{n+1,l}^{+} - 
\frac{1}{2} \hat{b}_{n,l}^{-}
\ee
and orthonormality to find
\bea
\bra{nl} r^2 \ket{nl} & = & 2n + l + \frac{3}{2} \label{quantumvirial} \\
\bra{n+1,l} r^2 \ket{nl} & = & \bra{nl} r^2 
\ket{n+1,l} \nonumber \\ & = & -\frac{1}{2} \sqrt{2 (n+1)(2 n + 2 l + 
3)}, 
\eea
with all other equal-$l$ matrix elements vanishing. (We recognize Eq. 
(\ref{quantumvirial}) as a consequence of the quantum virial theorem for the 
isotropic harmonic oscillator.)

For $l' = l+2$ (but $l \neq 0$), we begin with
\be
\bra{n l} r^2 \ket{n', l+2} =
\frac{\bra{n l} r^2 \hat{\cal L}_{l+1}^{+}  \hat{\cal L}_{l}^{+} 
\ket{n'+2, l}}{4 \sqrt{(n'+1)(n'+2)}}
\ee
and use the operator identity
 \bea
  r^2 \hat{\cal L}_{l+1}^{+} \hat{\cal L}_{l}^{+} &  = & \left[ 
  -\hat{b}_{n+1,l}^{-}  +  2 (n+1) + 2 \right] \times \nonumber \\
 &  & \left[ -\hat{b}_{n+2,l}^{-} + 2(n+2) \right]
 \eea
to obtain the non-vanishing matrix elements
\bea
\bra{n l} r^2  \ket{n, l+2} & = & \frac{1}{2} \sqrt{(2 n + 2l + 3)(2 n + 2 l +5)} \\
\bra{n+1, l} r^2 \ket{n, l & + & 2}  =  -\sqrt{2(n+1)(2 n + 2l + 5)} \\
\bra{n+2, l} r^2 \ket{n, l & + & 2}  =  \sqrt{(n+1)(n+2)}.
\eea

For the special case of $l=0$, we use the 
expansion of the irregular solutions in terms of 
$\ket{nl}$ \cite{Busch98a} and apply the above matrix elements. This 
results in
\bea
\bra{n0} r^2 \ket{n'2}   = 
\sqrt{\frac{2 \Gamma( n'+\frac{7}{2})}{\pi \Gamma(n'+1)}} a 
\sqrt{\fd{\nu_n}{a}} & \times & \nonumber \\
 \left(  \frac{1}{n'-\nu_n}  -   \frac{2}  {n'+1 - \nu_n} + 
 \frac{1}{n'+2 - \nu_n} 
\right) & . & 
\eea
(The matrix elements $\bra{n 0} r^2 \ket{n' 0}$ are not needed 
since $I_{00} = 0$.)

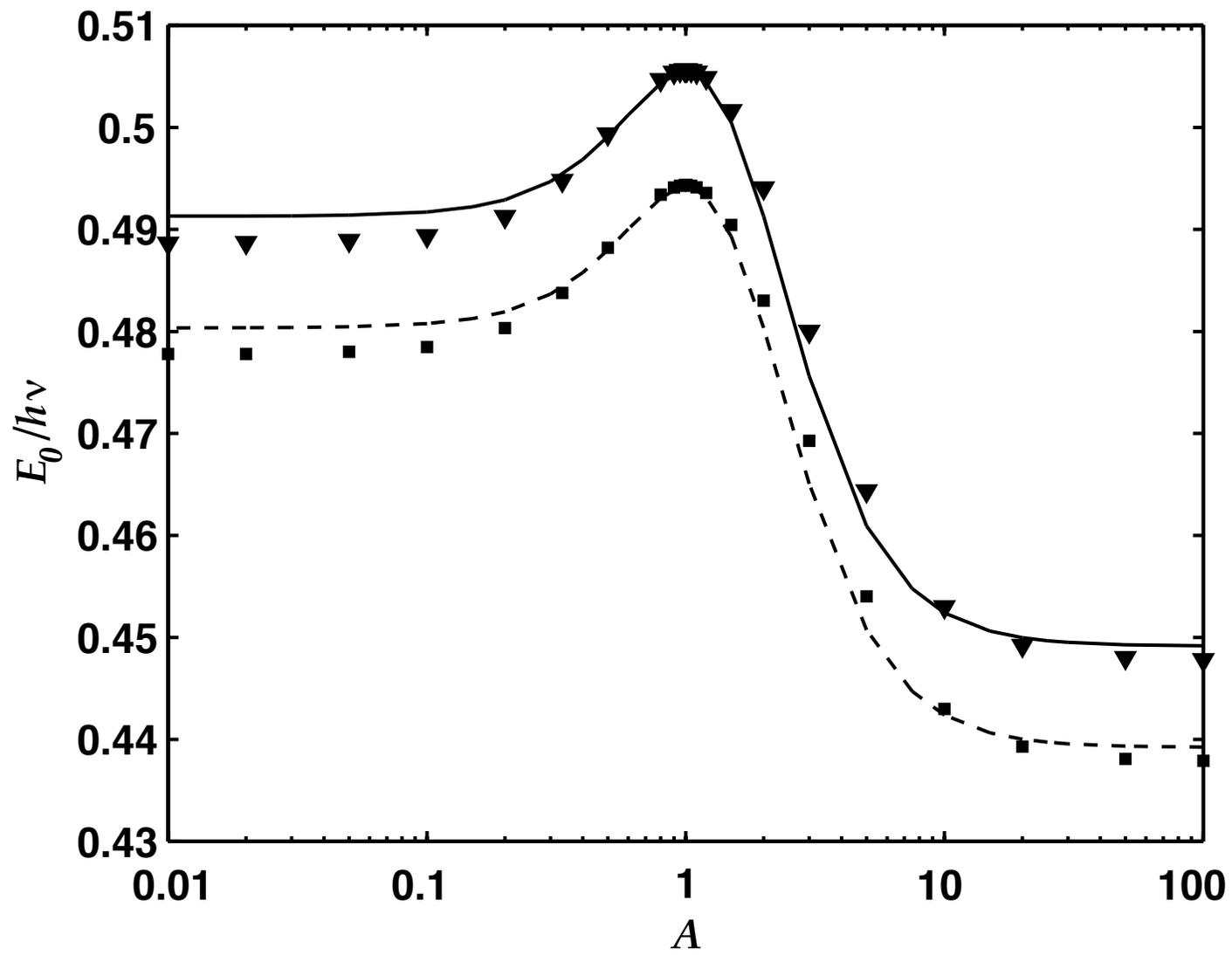
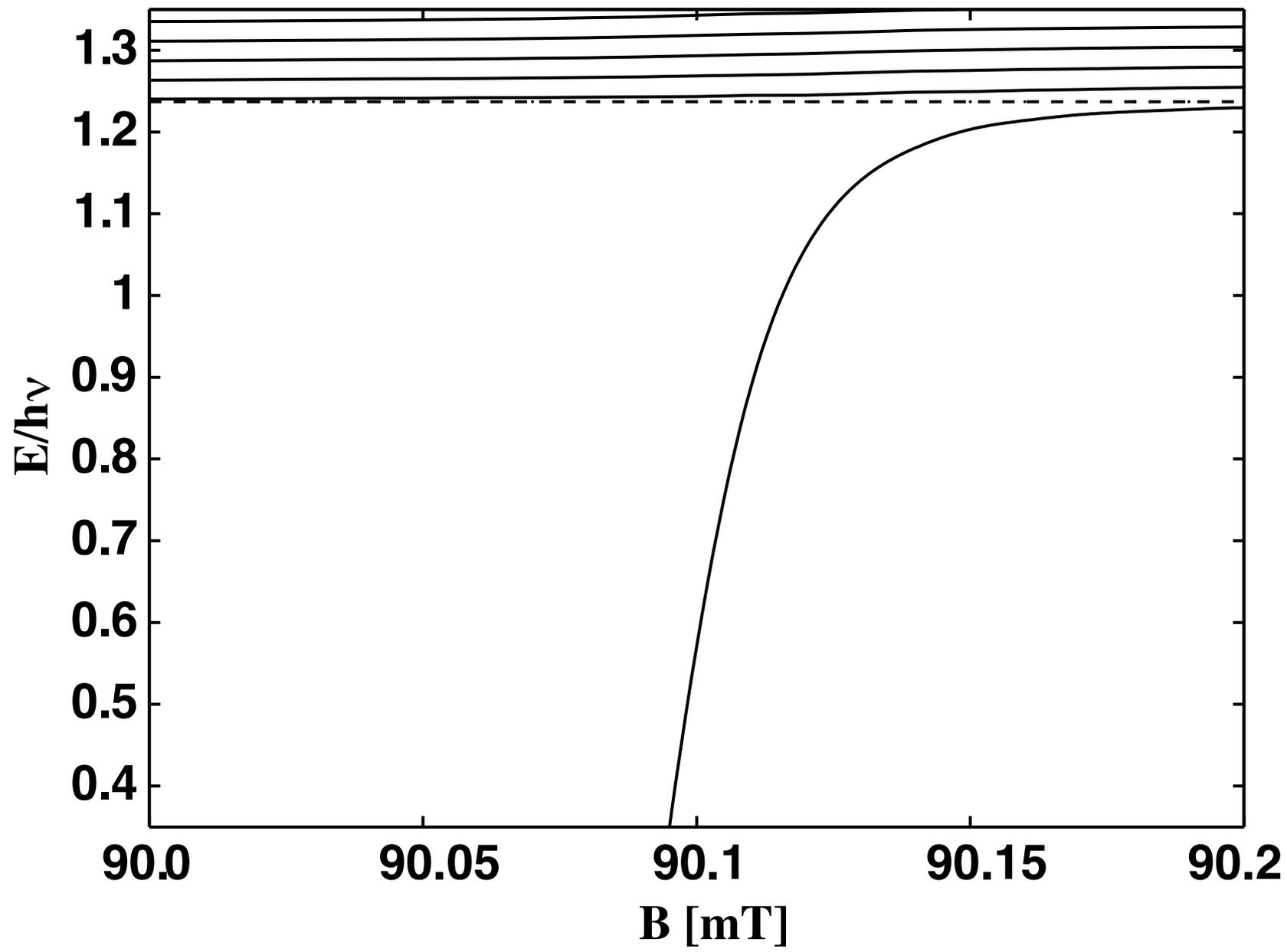
\end{document}